# Scalability and Performance Evaluation of Federated Learning Frameworks: A Comparative Analysis

**Bassel Soudan**
Department of Computer Engineering
College of Computing and Informatics
University of Sharjah
Sharjah, UAE
bsoudan@sharjah.ac.ae

**Sohail Abbas**
Department of Computer Science
College of Computing and Informatics
University of Sharjah
Sharjah, UAE
sabbas@sharjah.ac.ae

**Ahmed Kubba**
Department of Computer Science
College of Computing and Informatics
University of Sharjah
Sharjah, UAE
u23103280@sharjah.ac.ae

**Manar Wasif Abu Talib**
Department of Computer Science
College of Computing and Informatics
University of Sharjah
Sharjah, UAE
mtalib@sharjah.ac.ae
**Corresponding Author**

**Qassim Nasir**
Department of Computer Engineering
College of Computing and Informatics
University of Sharjah
Sharjah, UAE
nasir@sharjah.ac.ae

# 1 Abstract

This paper presents a systematic examination and experimental comparison of the prominent Federated Learning (FL) frameworks FedML, Flower, Substra, and OpenFL. The frameworks are evaluated experimentally by implementing Federated Learning over a varying number of clients, emphasizing a thorough analysis of scalability and key performance metrics. The study assesses the impact of increasing client counts on total training time, loss and accuracy values, and CPU and RAM usage. Results indicate distinct performance characteristics among the frameworks, with Flower displaying an unusually high loss, FedML achieving a notably low accuracy range of 66% to 79%, and Substra demonstrating good resource efficiency, albeit with an exponential growth in total training time. Notably, OpenFL emerges as the most scalable platform, demonstrating consistent accuracy, loss, and training time across different client counts. OpenFL's stable CPU and RAM underscore its reliability in real-world scenarios. This comprehensive analysis provides valuable insights into the relative performance of FL frameworks, offering good understanding of their capabilities and providing guidance for their effective deployment across diverse user bases.

# 3 Introduction

The traditional method for training machine learning (ML) models requires the collection of extensive datasets that represent the knowledge base of the problem at hand for the model to examine and learn. The ML model needs to have access to the raw data in the dataset with a significant set of attributes and descriptors that enable the categorization and classification of the data. This, however, poses a number of significant concerns regarding privacy and security. First, the data needs to be collected in one location (physical or virtual) to build the dataset. Any breach of this location will expose the entirety of the data to malicious actors. Additionally, the data may be sensitive, confidential, or otherwise controlled. Sharing it in its raw form will pose significant concerns.

Federated Learning (FL) was introduced in 2017 as a collaborative ML approach that operates without the need for storing the training data in one location [1]. It offers a secure approach where collaborating clients train a ML model without the need for sharing sensitive data. Each client trains a private replica of the model locally using its own private data. Then, only the trained *model parameters* are shared with a global server that performs aggregation and redistribution to all clients. After the updates are received from the server, each client updates the parameters of its local model, and the process repeats. No training data is ever shared amongst the clients, and each client maintains complete control over its private data.

Federated Learning transfers the responsibility of model training to the individual clients, with communication between clients and the server occurring through parameter interaction instead of direct data interaction. The server's role is limited to simple parameter aggregation for updating the global model as shown in Figure 1. This scheme enables the protection of local user data while conserving the server's compute and storage resources [2].

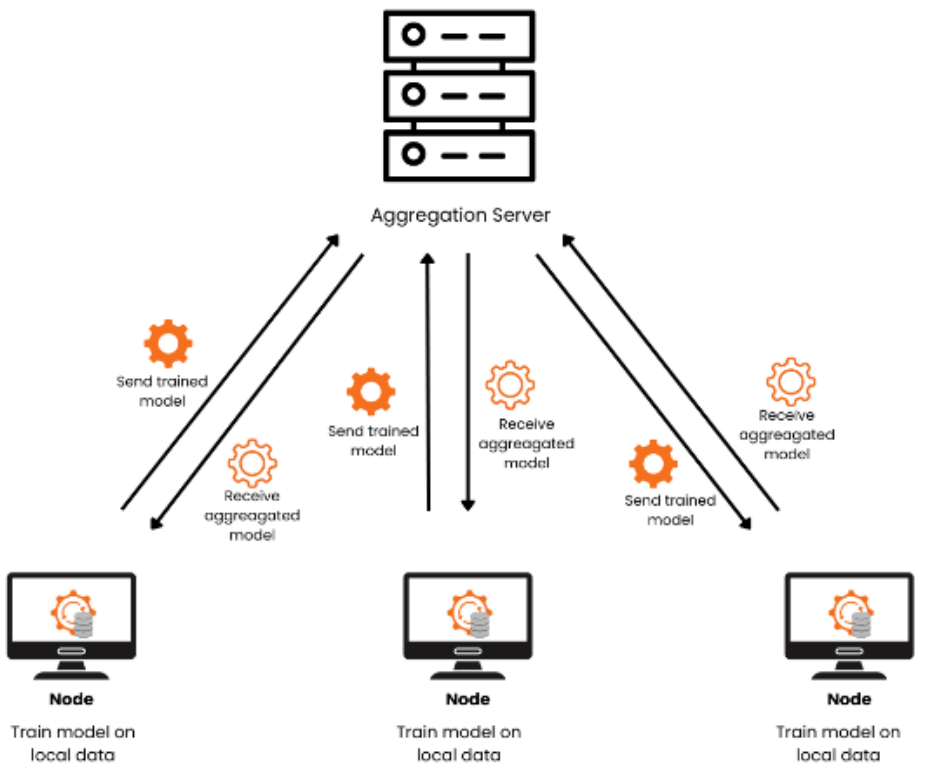

Figure 1 – Typical framework of a Federated Learning system

Federated Learning employs the principle of "data minimization" which means that during training in a FL system, only minimal updates for a specific model training task are transmitted [3]. Access to data at every stage is restricted, individual data is processed as early as possible, and processed and collected data are discarded as quickly as possible.

Several frameworks have been developed through research and private enterprise to facilitate the implementation of FL in various applications. The following are the most commonly used frameworks, at the time of this writing:

1. **Flower (Federated Learning Framework).** Flower is an open-source Python library that allows model training across decentralized devices or servers. Flower works with popular deep learning frameworks like TensorFlow and PyTorch. It is designed to scale to a large number of participating clients and supports diverse FL device scenarios. Flower follows the client-server architecture where the central server manages the global model, and the clients perform local training. It supports optimizations for enhancing the efficiency and speed of the FL processes [4].
2. **The FedML platform.** FedML is an open research library and benchmark that supports diverse FL computing paradigms: on-device training for edge devices, distributed computing, and single-machine simulation [5]. FedML also supports flexible and generic API design and provides comprehensive reference baseline implementations (optimizer, models, and datasets). FedML utilizes a "virtual nodes and central server" architecture [6].
3. **Substra.** Substra is an open source framework which provides distributed learning that guarantees privacy such that the data never leaves the nodes, and the exchange within the network involves only the predictive models, algorithms, and non-sensitive metadata. It supports a variety of computation models, including the parallel computation plan used in FL [7].

4. **The Open Federated Learning (OpenFL) framework.** OpenFL is an open-source Python-based tool designed for training machine learning algorithms through the collaborative learning paradigm of FL. It is compatible with training pipelines using both TensorFlow and PyTorch and can be extended to other ML frameworks [8]. The significance of OpenFL lies in its scalability, emphasis on trusted execution, and the consistent migration of centralized ML models to a Federated Training pipeline [9].

These frameworks offer a range of different features, advantages, and disadvantages that affect FL's adoption in various applications. To make an informed decision when it comes to selecting a certain FL framework for a particular application, it is important to understand which framework aligns best with the application's requirements. That is why a comprehensive analysis of these frameworks, which focuses on their capabilities and performance, is essential to identify the most suitable choice for a given application.

A comparative analysis of open-source FL frameworks has been conducted with the aim of assessing their suitability for IoT systems [10]. The evaluation considered factors such as ease of deployment, development, analysis capabilities, accuracy, and performance. The study employed three datasets and modeled low-power IoT devices with limited computing resources. The analysis concluded with the identification of FL frameworks that can be utilized under the resource restriction of IoT systems, although with certain usage restrictions.

Another work presented a comprehensive examination of FL with the aim of offering a thorough summary of relevant FL protocols, platforms, and real-life use-cases [11]. The work explored both the challenges and advantages of FL, presenting detailed service use-cases to illustrate how various architectures and protocols utilizing FL can synergize to achieve desired outcomes.

An assessment of the performance, scalability, and utilization of the Flower, FedN, and FedML frameworks was presented using a Natural Language Processing (NLP) use case [6]. The work presented a detailed examination of the frameworks covering their structure, communication methods, and integration with DL libraries. Performance was evaluated through experiments on a standard benchmark dataset, measuring accuracy, speed, and scalability.

An assessment of the appropriateness of existing FL libraries for industrial applications has been conducted [12]. The work presented a comparative analysis based on review of the scientific literature and official documentation of the different frameworks. The work also presented a benchmarking tool to assess the non-functional characteristics of the libraries.

It is clear from this quick survey of existing works that none of the works presents a comprehensive evaluation of all of the frameworks. Also, none of these previous works has performed benchmarking that identifies the scalability and expansion of the different frameworks. Finally, none of these works seems to address the resource efficiency of the frameworks. Accordingly, the primary objective of this research is to conduct a comprehensive and systematic examination of the prominent FL frameworks: FedML, Flower, Substra, and OpenFL. Through experimental comparisons, this study aims to assess the scalability and key performance metrics of these frameworks, emphasizing the impact of varying client counts on total training time, loss and accuracy values, as well as CPU and RAM usage. The predominant goal is to provide a thorough comparative analysis that identifies the distinct capabilities, strengths, and weaknesses of each FL framework. Additionally, this research endeavors to establish a benchmark, serving as a reliable reference point for evaluating the effectiveness of these frameworks in diverse scenarios. The contributions of this work can be expressed as follows:

1. Conduct a detailed comparative analysis of the selected Federated Learning frameworks to evaluate their capabilities.
2. Identify and highlight the strengths and weaknesses of each FL framework.
3. Establish a benchmark that provides a reliable reference point for assessing the effectiveness of the frameworks.
4. Assess the resource efficiency of each framework, focusing on computational resource utilization.

# 4 Background

Federated Learning is a collaborative approach that allows training ML models without the need to share sensitive training data. FL clients collaborate in training replicas of the model locally on their respective hardware, using their respective data. After each training round the clients communicate *only* the model parameters to a global server. The server aggregates the parameters from all of the clients to generate a *global parameter update*. This global update is then communicated back to the individual clients who in turn update the parameters of their local models. The process continues until the desired outcome is reached and the local models on the clients are fully trained. The FL model operates without the need to share sensitive user data with any other participant in the system. Actually, the user data never leaves the security of the client on which it resides.

To enhance the security further, even the model parameters are communicated in an encrypted manner between the clients and the global server. A Secure Aggregation Protocol is employed to aggregate the client updates into the global update [13]. This secure aggregation protocol enables the coordinating server to decrypt the client updates only when hundreds or thousands of clients have participated. Consequently, individual updates from specific devices cannot be examined prior to aggregation. This aggregation protocol is specifically designed to handle deep-network-sized problems and account for real-world connectivity limitations.

A considerable challenge for the implementation of FL systems is the wide diversity in the communication and computation capabilities of devices within the federated network. This diversity poses challenges for maintaining fault tolerance and ensuring consistent progress amongst the disparate clients. It is also crucial to ensure that the FL system can accommodate devices that become temporarily inactive during the training process for the model [14].

## 4.1 Categorization of Federated Learning based on Data Distribution

Given that the model training and the training dataset are distributed between the different clients, this creates a number of variations in terms of how the training is carried out (as demonstrated in the depictions of Figure 2) [15]. The following is a brief introduction to the different flavors:

1. Horizontal Federated Learning - Horizontal Federated Learning (HFL) involves the collaboration of different entities that possess similar types of data but distinct samples [1], [16]. An example would be different healthcare providers that collectively leverage their individual datasets that have similar features, but different patients. HFL allows these entities to collaboratively train a shared ML model on the combined population of patients without compromising the confidentiality of their individual datasets. HFL allows expanding the size of the user sample without the need for the collaborating entities to share their private sensitive data.
2. Vertical Federated Learning - Vertical Federated Learning (VFL) involves the collaboration of different entities possessing complementary but non-overlapping sets of data features [1], [17]. In this scenario, the collaborating parties typically share a *common set of instances*, but *different features*. A practical example would be two organizations possessing data on the same set of individuals but with distinct attributes, such as different health records or financial information. VFL allows expanding the feature dimension for the similar instances. This approach allows for valuable insights to be gained from the combination of diverse data attributes, fostering collaborative knowledge extraction without compromising individual data security.
3. Federated Transfer Learning - Federated Transfer Learning (FTL) allows multiple decentralized entities to collaborate by leveraging the knowledge gained from locally collected data to enhance the learning capabilities of a shared model [18]–[20]. FTL allows the model to transfer knowledge learned from one domain to improve performance in another. For example, if one dataset comes from an e-commerce company in China, and the other from a social application in the United States. The user populations in these datasets barely intersect due to geographical limits. Furthermore, their data features have limited overlap due to the different nature of the institutions. FTL is especially useful when it is needed to optimize the performance of a task but there’s a lack of sufficient related data for training. It helps address challenges like data scarcity and data heterogeneity.

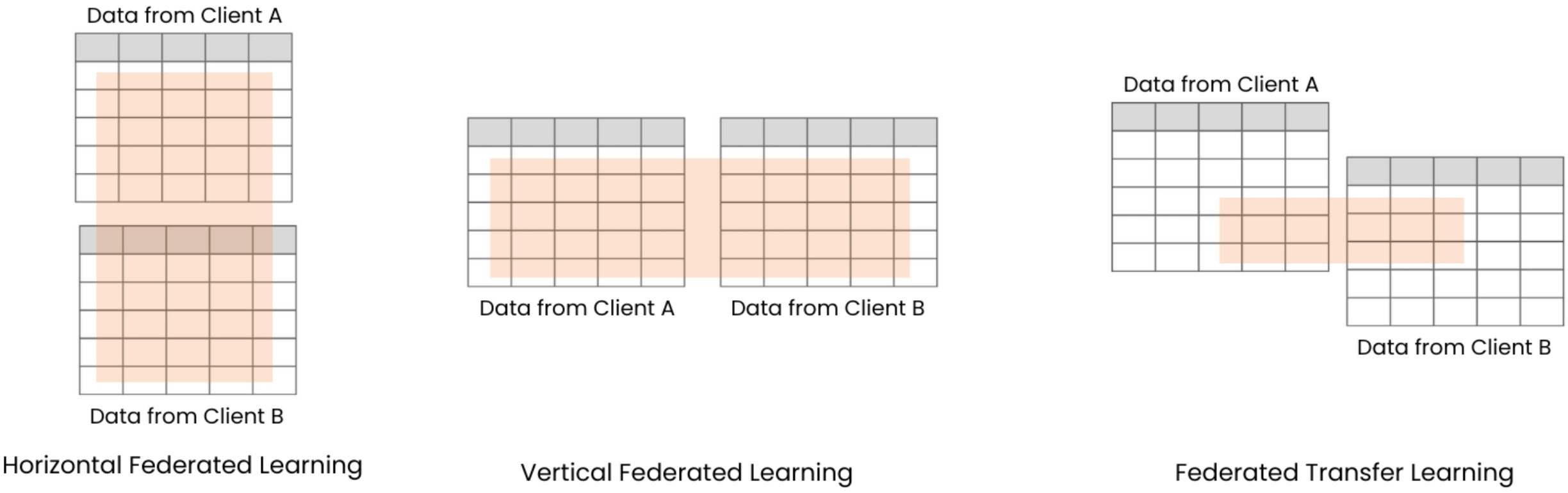

Figure 2 – Types of Federated Learning Based on Distribution of the Data in the Local Datasets

## 4.2 Categorization of Federated Learning based on Type of Participating Client

Federated Learning can be categorized into two categories based on the capabilities, count, and type of participating clients: cross-device FL and cross-silo FL (as demonstrated in the diagrams of Figure 3). In cross-device FL, the participating clients are compact distributed entities such as smartphones, wearables, and edge devices [21], [22]. Each client generally possesses a relatively limited amount of local data. Consequently, the success of cross-device FL often hinges on the involvement of a substantial number (potentially millions) of edge devices in the training process. On the other hand, cross-silo FL involves clients that are typically companies or organizations like hospitals and banks [23], [24]. In this scenario, the number of participants is modest, ranging from two to a hundred, and each client is expected to engage in the entire training process.

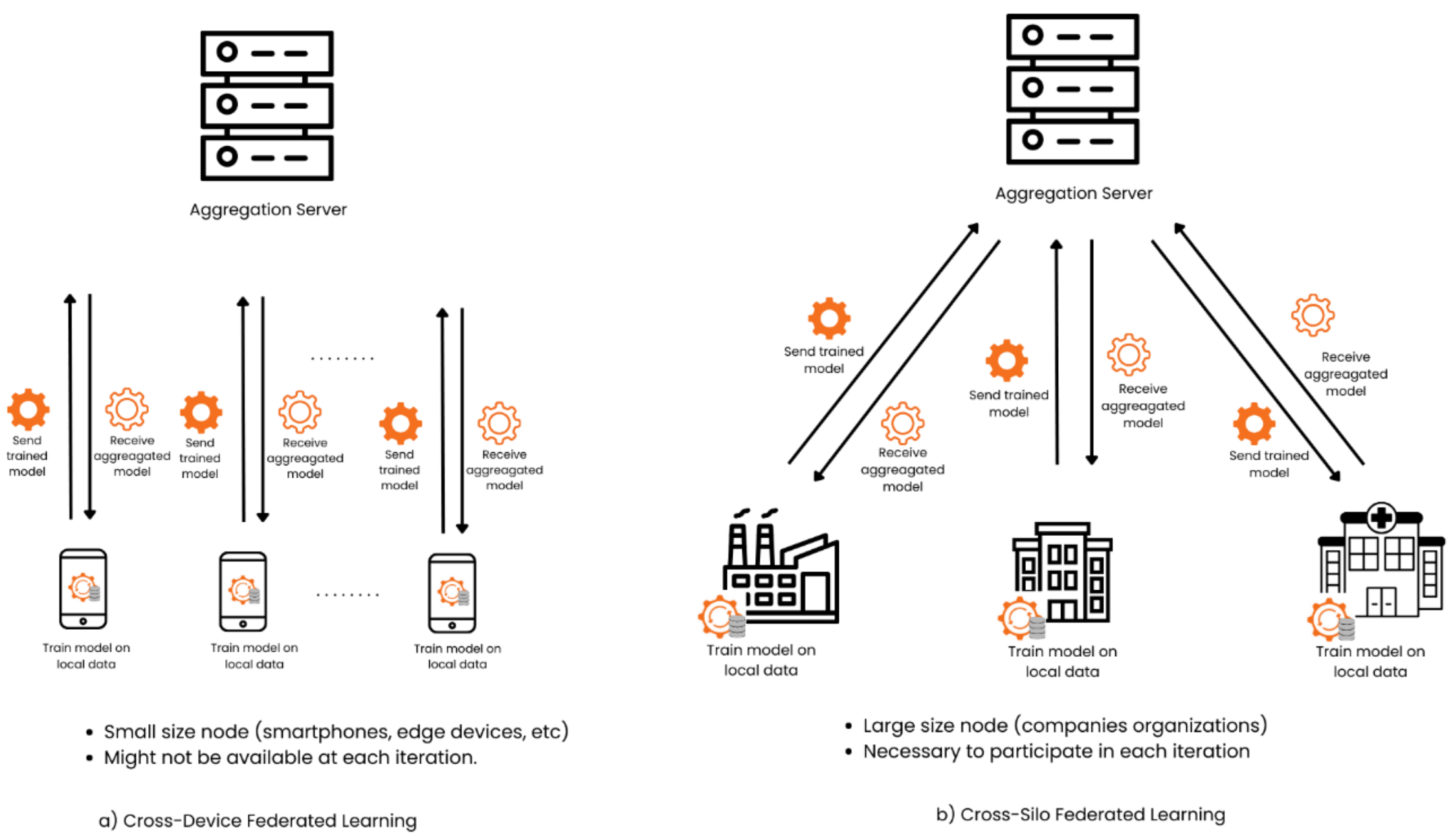

Figure 3 –Cross-Device versus Cross-Silo Federated Learning

## 4.3 Security in Federate Learning

Federated Learning can employ a variety of security approaches to further ensure secure aggregation and privacy preservation during the federated training process.

- **Secure Aggregation** ensures that individual updates from clients are aggregated in a way that the server can only access the aggregated results and not the individual contributions. An example of secure aggregation is the use of cryptographic techniques similar to Homomorphic Encryption to safeguard model updates during aggregation, preventing unauthorized access [25]. Another example is Secure Multi-Party Computation (SMPC) where the computations are performed on encrypted data, ensuring that no party has access to the raw data [26].
- **Differential Privacy** introduces controlled noise into the updates, making it difficult to infer any individual's data from the shared information [25]. In the context of Federated Learning, each client has the ability to inject randomly generated values into its model weights prior to transmission. As a result, even if the data is subjected to reverse engineering, it does not correspond precisely to the data of any individual user.
- **Byzantine Fault-Tolerant Aggregation (BFT)** is an approach for tackling Byzantine faults where a subset of the nodes become compromised or start to behave maliciously [27]. The primary challenge posed by Byzantine faults is their unpredictability and the potential for significant disruption or corruption of the learning process. The goal of BFT aggregation is to enable a system to continue functioning correctly even in the presence of a fraction of faulty or malicious nodes. This involves the use of algorithms and techniques that can detect, isolate, or mitigate the influence of these Byzantine faults during the aggregation process [28].
- **Anomaly Detection** focuses on identifying patterns or behaviors that deviate from the expected norm. In the context of FL, anomalies may arise from compromised nodes, adversarial attacks, or unexpected hardware/software behaviors. The goal of anomaly detection is to identify and mitigate these anomalies promptly to safeguard the integrity of the model [29].

It is important to consider support for these aspects of the Federated Learning process when determining which is the most appropriate framework to use for a given application. The following section will introduce the methodology that was used for evaluating the different frameworks mentioned in Section 3.

# 5 Methodology

A well-structured and thorough approach has been adopted in the evaluation of the FL frameworks to ensure credibility and precision. This section describes the methodology employed for the evaluation, encompassing the chosen frameworks, the experimental dataset, and the metrics used for assessment. The proposed methodology aims to establish a systematic and thorough foundation for the ensuing comparative analyses.

## 5.1 Selection of FL Frameworks

The selection process for FL frameworks to be evaluated focused on open-source frameworks that represent a broad spectrum of use-cases and distinct operational requirements. The frameworks chosen for this study (Flower, FedML, Substra, and OpenFL) are among the most frequently cited in academic and industry literature, indicating their prominence and widespread adoption. As illustrated in Table 1, these frameworks demonstrate a diverse range of features and capabilities, providing a comprehensive overview of the current landscape of FL tools and technologies.

Table 1 – Features and Capabilities of the Selected Federated Learning Platforms

| Features | | Flower | FedML | Substra | OpenFL |
|---|---|---|---|---|---|
| Operating Systems | MacOS | ✓ | ✓ | ✓ | - |
| | Linux | ✓ | ✓ | ✓ | ✓ |
| | Windows | ✓ | ✓ | ✓ | ✓ |
| | iOS | ✓ | ✓ | - | - |
| | Android | ✓ | ✓ | - | - |
| Data Partitioning | Vertical Split | ✓ | ✓ | - | - |
| | Horizontal Split | ✓ | ✓ | ✓ | ✓ |
| Deployment Scenarios | Cross-Silo | ✓ | ✓ | ✓ | ✓ |
| | Cross-Device | ✓ | ✓ | - | - |
| Privacy Mechanism | Differential Privacy | ✓ | ✓ | ✓ | ✓ |
| | Cryptographic Techniques | ✓ | ✓ | ✓ | ✓ |
| Model Support | Model-agnostic | ✓ | - | ✓ | ✓* |
| | Limited | - | ✓ | - | - |

*OpenFL can support additional ML models through an extensible mechanism [9].

## 5.2 Dataset Selection

A number of experiments have been undertaken, employing the widely recognized MNIST dataset, with the primary objective of establishing a performance benchmark for various frameworks [30]. This dataset, a well-known and extensively studied resource in machine learning research, was chosen for evaluation due to its simplicity, scalability, and resource efficiency. Comprising a substantial collection of images depicting hand-written digits, the MNIST dataset encompasses 60,000 training images and 10,000 testing images, providing a robust foundation for assessing the capabilities of the frameworks.

## 5.3 Experimental Setup

Simulated training environments were set up for each framework to simulate the process of FL across many clients, using the chosen dataset and aggregation algorithm. All experiments were conducted on a workstation running Windows 10 with 32 GB of RAM and an Nvidia RTX 4000 Graphical Processing Unit. The Central Processing Unit of the workstation is Intel® Xeon® W-2102 with four cores and a base speed of 2.90 GHz.

All clients and the aggregation server are executed in parallel on the same workstation utilizing multi-threading across multiple cores. The execution times for all operations are measured using built-in time measurement functions, encompassing the entire spectrum of activities related to both client and server tasks in the training process. The tool used to record the training time is Psutil, which is a cross-platform python library that provides data on running processes and system usage on the workstation [31]. These measurements also include any delays associated with queuing and messaging between client and aggregation server processes. Given that all processes are executed on the same workstation; no consideration was given for network communication delays. It is important to highlight that this omission does not compromise the accuracy of the framework's performance evaluation. Communication delays are more a measure of the performance of the network rather than the framework itself. Thus, the focus remains on assessing the framework's intrinsic capabilities without introducing network-related variables into the evaluation process.

It is vital to simulate the real-world scenario where each client possesses its own dataset, even though it was decided to use a standard dataset from previous research work. Therefore, the MNIST dataset will be split according to the requirements of HFL and VFL. Then, the individual sub-datasets will be distributed to the different clients before the training process is commenced. The time and effort needed for this operation will be excluded from the evaluations of the framework as it is not related to the framework's performance.

## 5.4 Evaluation Metrics

A number of relevant evaluation metrics have been selected to measure the performance of the different frameworks. The chosen metrics are:

1. **Loss and accuracy values:** These metrics serve to gauge the effectiveness of the ML training process in the different frameworks.
2. **Total training time:** This metric provides insights into the responsiveness of the framework, offering a measure of its efficiency in completing the training tasks.
3. **CPU and RAM usage during training:** These metrics are essential for determining the framework's efficiency and suitability under varying resource constraints, providing a comprehensive understanding of its resource utilization patterns. This metric was recorded using the psutil python tool [31].

# 6 Results and Discussion

This section presents a thorough examination of the results and outcomes of the evaluation experiments. The section aims to shed light on the performance, intricacies, and comparative aspects of the frameworks identified in Section 5.10. The frameworks were evaluated both quantitatively (using the evaluation metrics mentioned in Section 5.4) as well as qualitatively where the different features and capabilities of the frameworks were compared. Therefore, the quantitative and qualitative results will be presented separately.

## 6.1 Quantitative Performance Evaluation of the Different Frameworks

A large number of experiments were conducted to evaluate the performance of the different frameworks. The experiments evaluated the scalability of the framework by varying the number of clients from 20 to 100 clients. Each experiment was executed for a constant 20 federated training rounds, with each round spanning 15 local epochs. The experiments are repeated as necessary to ensure reproducibility and statistical significance. The results presented in Table 2 present the average of the performance metrics for the different repetitions of each experiment.

Table 2 – Performance Metrics for the Different Frameworks with Different Client Counts

| Framework | Number of Clients | Total Training Time (Sec) | Loss | Accuracy | CPU Usage % | RAM Usage % |
|---|---|---|---|---|---|---|
| Flower | 2 | 4192 | 4.582 | 0.9931 | 80.08 | 45.43 |
| | 15 | 6008 | 2.742 | 0.9913 | 76.14 | 47.92 |
| | 30 | 5712 | 3.642 | 0.9891 | 75.69 | 47.85 |
| | 60 | 5044 | 3.906 | 0.9874 | 89.65 | 52.11 |
| | 80 | 4248 | 5.298 | 0.9835 | 86.61 | 46.03 |
| | 100 | 4539 | 5.314 | 0.9809 | 85.93 | 49.02 |
| FedML | 2 | 271 | 1.109 | 0.664 | 90.48 | 30.49 |
| | 15 | 707 | 1.038 | 0.6579 | 93.45 | 30.28 |
| | 30 | 1,431 | 1.869 | 0.5212 | 96.85 | 30.43 |
| | 60 | 2,633 | 1.035 | 0.6819 | 98.24 | 30.62 |
| | 80 | 3,269 | 0.687 | 0.7662 | 98.62 | 31.14 |
| | 100 | 4,346 | 0.615 | 0.7913 | 98.97 | 31.58 |
| Substra | 2 | 180 | 0.278 | 0.9148 | 42.18 | 20.71 |
| | 15 | 1330 | 0.166 | 0.9537 | 42.88 | 21.05 |
| | 30 | 2,760 | 0.243 | 0.9098 | 42.15 | 20.64 |
| | 60 | 6,320 | 0.194 | 0.9467 | 40.73 | 21.27 |
| | 80 | 9,590 | 0.239 | 0.9365 | 38.46 | 23.21 |
| | 100 | 13,630 | 0.202 | 0.9406 | 35.81 | 21.38 |
| OpenFL | 2 | 2246 | 0.35 | 0.9743 | 67.19 | 42.76 |
| | 15 | 2816 | 0.122 | 0.9738 | 65.26 | 42.29 |
| | 30 | 3128 | 0.145 | 0.9674 | 64.99 | 45.89 |
| | 60 | 4269 | 0.169 | 0.9594 | 63.1 | 53.13 |
| | 80 | 4985 | 0.177 | 0.9567 | 62.14 | 58.34 |
| | 100 | 5876 | 0.193 | 0.949 | 64.13 | 63.11 |

The comparative results of the different experiments for each performance metric are presented in Figure 4 through Figure 8. These results will now be discussed first based on the individual evaluation metrics, then holistically for the individual frameworks.

### 6.1.1 Loss and Accuracy

The graphs in Figure 4 and Figure 5 show the loss and accuracy measurements for the different frameworks using different client counts.

The data in Figure 4 illustrate that on the whole, OpenFL demonstrates the minimal loss with an average of only 0.192, followed closely by Substra with an average loss of 0.22, while FedML exhibits an average loss of 1.058. All of which are notably lower than Flower, which records an average loss of 4.247.

Additionally, the data show that the loss measurement is significantly affected by the number of clients for all of the frameworks, with a variability reaching as high as 67% for FedML, 65% for OpenFL, 48% for Flower, and 40% for Substra. Notably, FedML stands out as its loss seems to reduce consistently with the increase in the number of clients.

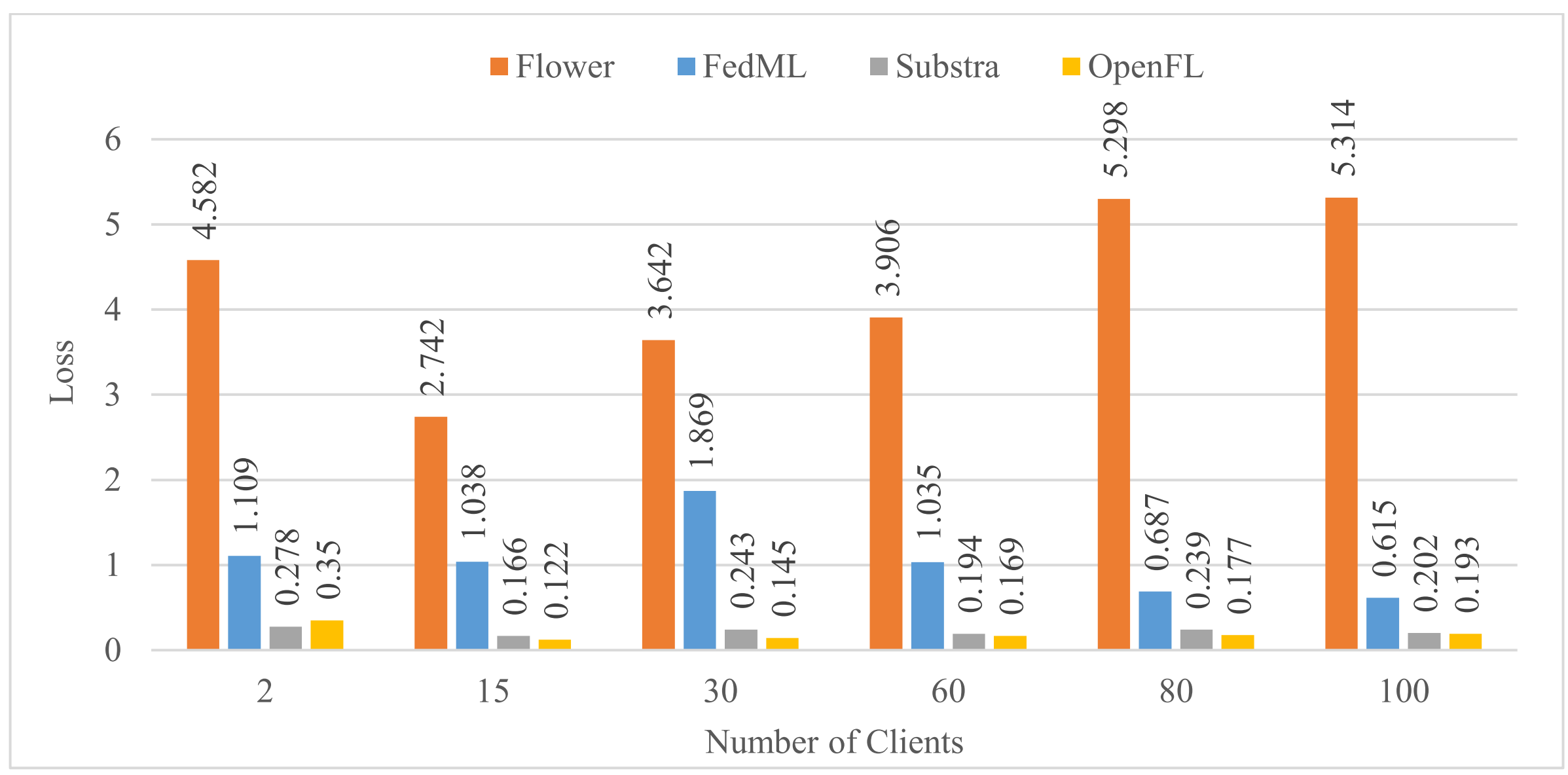

Figure 4 – Loss Measurement for the Different Frameworks with Different Client Counts

On the other hand, the data in Figure 5 demonstrate high levels of accuracy measurement for all frameworks, with the notable exception of FedML, which consistently scored significantly lower under all client counts. The Flower framework produced that highest consistent accuracy with an average of 98.75% and a mere 1% variation in accuracy with the different client counts. OpenFL and Substra exhibited reasonable accuracy with an average of 96.34% for OpenFL and 93.367% for Substra. Both frameworks also exhibited robust accuracy with less than 2% variability. FedML on the other hand, showed substantially reduced accuracy that reached only in the range of ranged from 66% to 79%. FedML also exhibited significant variability in accuracy with the different client counts.

In general, it is clear that OpenFL offers the best choice from the point of view of loss, with a caveat that the loss seems to increase with increasing client counts. On the other hand, it is clear that FedML has an issue with achieving adequate accuracy for the training process. Flower seems to present the most compelling choice from this point of view.

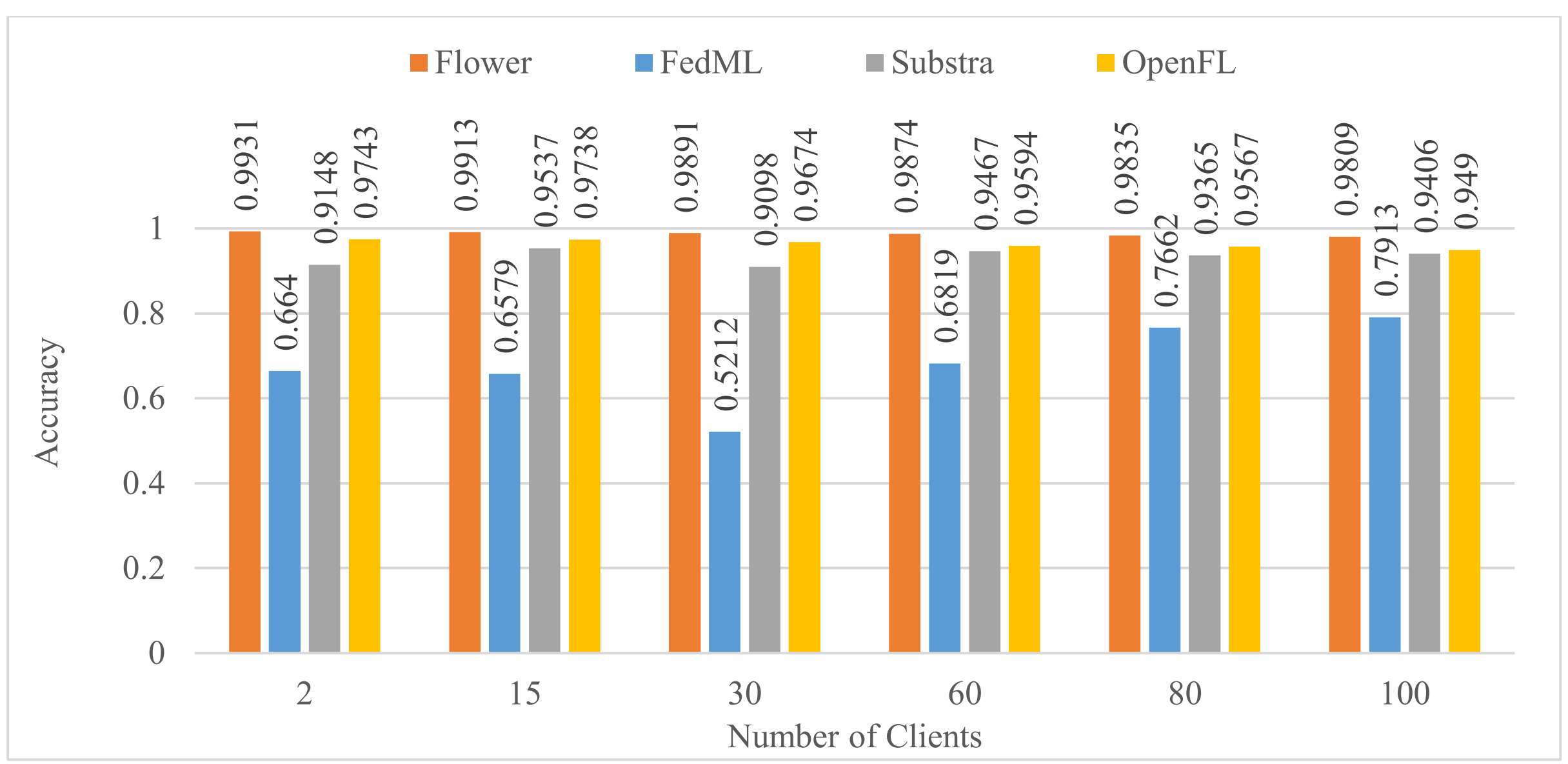

Figure 5 – Accuracy Measurement for the Different Frameworks with Different Client Counts

### 6.1.2 Total Training Time

The plot in Figure 6 presents the total training time for the different frameworks using the different client counts. It can be summarized from the figure that in general training time increases for all frameworks with the increasing number of clients. It is clear from the figure that Substra has an issue with scalability as its total training time seems to grow almost exponentially with the number of clients. On the other hand, total training time for FedML and OpenFL seems to grow linearly with the number of clients. The notable exception is Flower whose total training time seems to be reasonably consistent, albeit generally high compared to the other frameworks. It is clear that Flower has a constantly high training overhead, even for the small number of clients, that balances out the effect of increasing the client count.

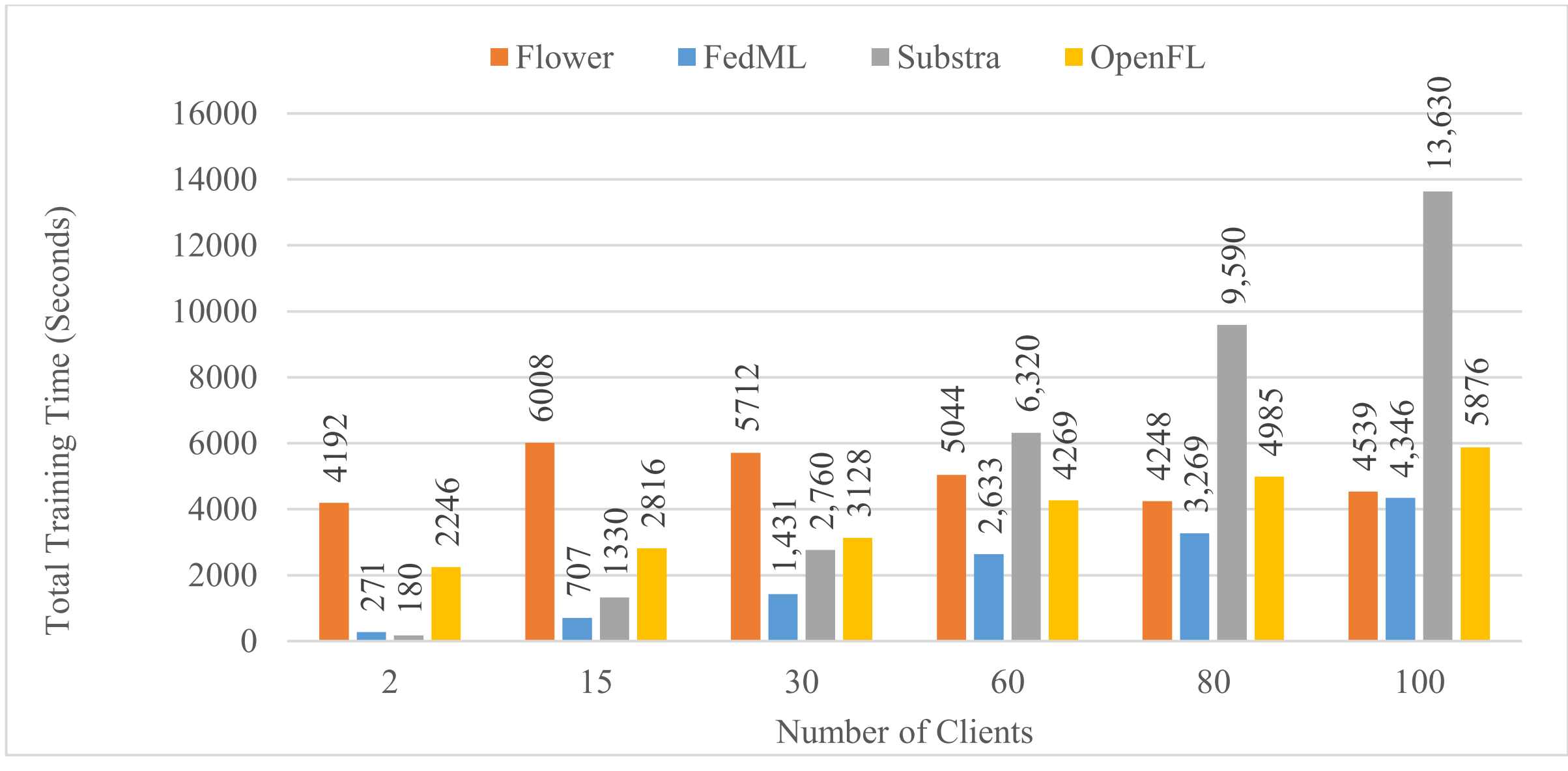

Figure 6 – Total Training Time for the Different Frameworks with Different Client Counts

In general, it seems that FedML and OpenFL both exhibit robust training time that has reasonable overhead which grows linearly with the number of clients.

### 6.1.3 Resource Usage Efficiency

The plots in Figure 7 and Figure 8 demonstrate the CPU and RAM usage for the different frameworks using different client counts. The data in Figure 7 show that Substra seems to have the most efficient CPU usage as it demonstrates a consistent average of 40% of CPU power being used to support the operations of all clients as well as the aggregation server. OpenFL is a close second with a consistent 64% average CPU usage. Flower exhibited an average higher than 80%, while FedML consumed the most CPU operations exceeding 96% of CPU cycles on average. Interestingly, CPU usage for Substra and OpenFL seemed to reduce with the increasing number of clients. This can probably be attributed to a higher percentage of communication overhead (which is discounted in our measurements) with the larger number of clients.

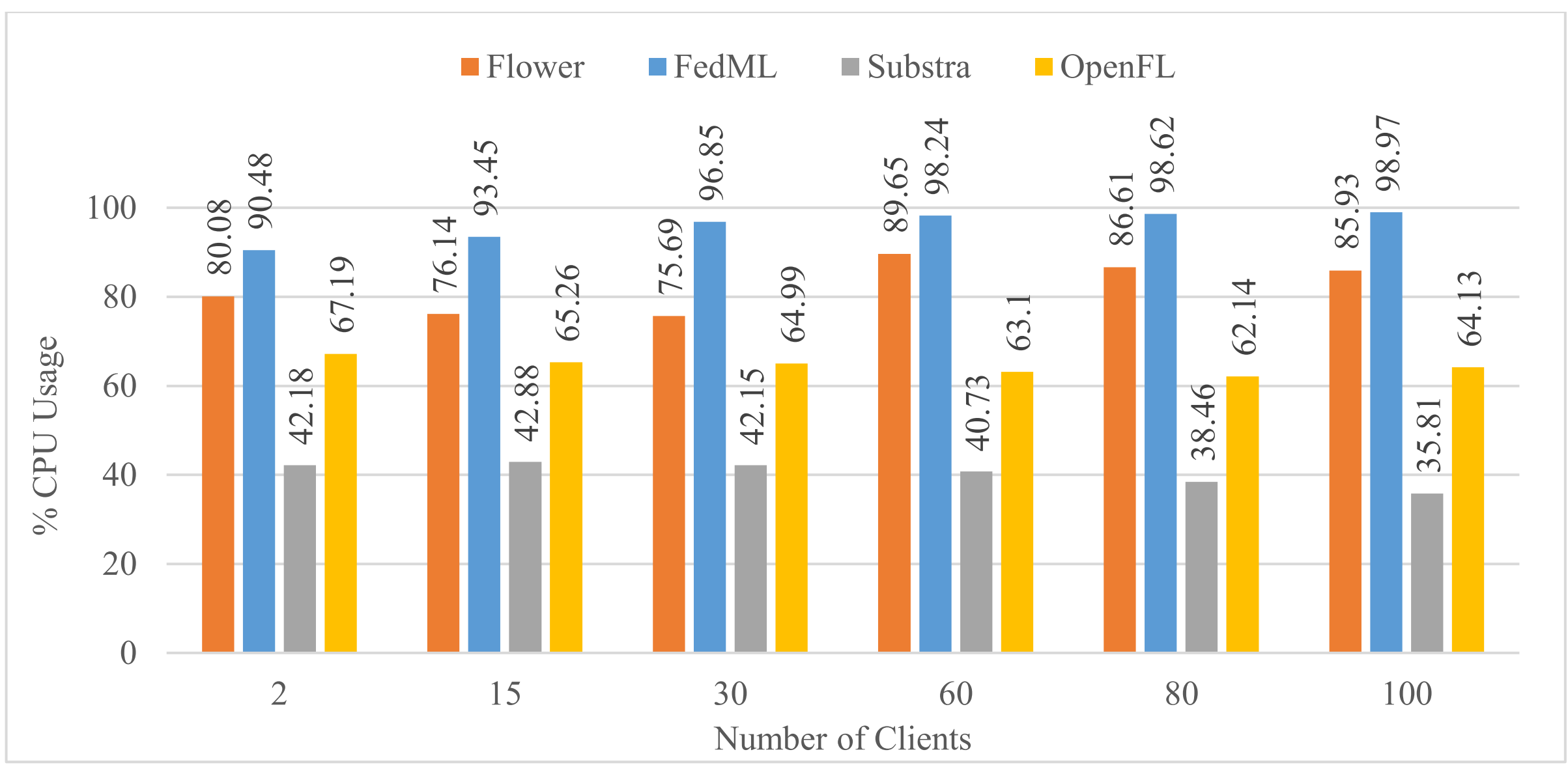

Figure 7 – Percent CPU Usage for the Different Frameworks with Different Client Counts

On the other hand, the plot in Figure 8 shows the memory usage to support the different frameworks with the different client counts. In general, all frameworks seem to have a reasonable level of memory occupancy given the complex operations being implemented. Substra again seems to be the most efficient with a consistent average of about 21% memory occupancy, closely followed by FedML, then Flower. OpenFL is the exception where memory occupancy increases significantly with the number of clients.

Based on resources utilization, it seems that Substra is the most efficient framework, and would probably lend itself better to resource-limited scenarios such IoT implementation. FedML has a very high computational requirement (as exhibited by the very high CPU usage). Therefore, FedML does not seem to be suitable for resource-limited applications.

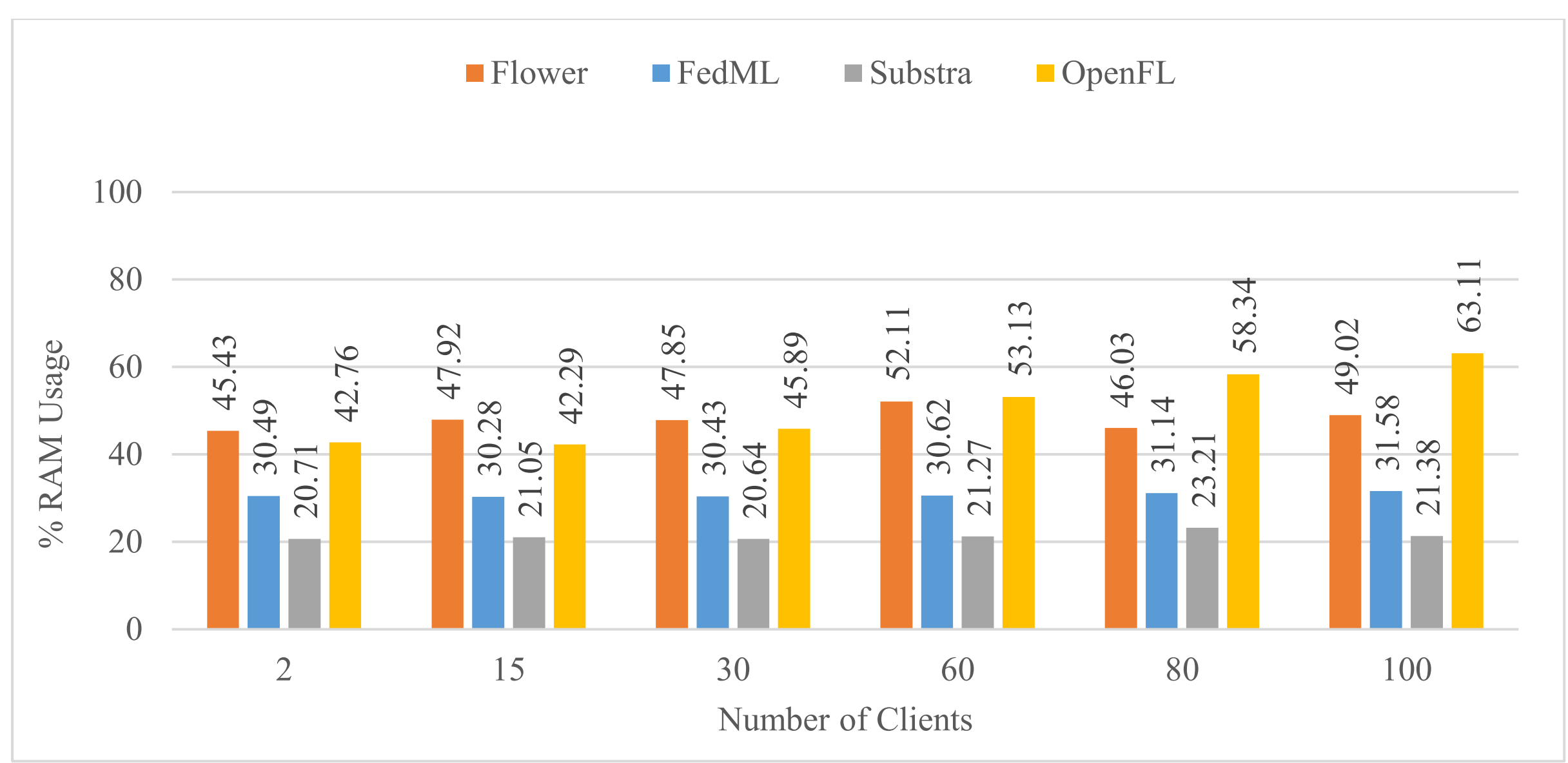

Figure 8 – Percent RAM Usage for the Different Frameworks with Different Client Counts

## 6.2 Holistic Evaluation of the Individual Frameworks

The observations in Table 3 summarize the findings of the quantitative evaluation of the different framework holistically.

Table 3 – Holistic Comparison of the Performance of the Different Frameworks

| Evaluation Metric | Framework | | | |
|---|---|---|---|---|
| | **Flower** | **FedML** | **Substra** | **OpenFL** |
| **Loss** | • Exceptionally high loss value compared to other frameworks<br>• Loss value increases with the increasing number of clients | • Loss value in the range of 1%<br>• Some variation in loss value with scaling client counts | • Very low loss value, in the range of 0.2%<br>• Very little variation in loss value. | • Best loss value, below 0.2%<br>• Insignificant variation in loss value with scaling client count. |
| **Accuracy** | • Highest accuracy, consistently exceeding 98%<br>• Accuracy not affected by scaling | • Exceptionally low accuracy, averaging only 68% | • Reasonable accuracy, in the range of 93%<br>• Accuracy not affected by scaling | • Very good accuracy, in the range of 96%<br>• Accuracy not affected by scaling |
| **Training Time** | • Consistently high training time, in the range of about 5000 seconds<br>• Training time not affected by scaling. | • Best training time trend.<br>• Training time increases linearly with scaling client count. | • Training time grows exponentially with scaling client count. | • Somewhat elevated training time that grows linearly with scaling client count. |
| **CPU Usage** | • High computation load that shows little variation with scaling client count | • Exceptionally high computation load, in the range of 99% CPU usage. | • Lowest computation load that reduces with scaling client count. | • Average computation load, with insignificant variation with scaling. |
| **RAM Usage** | • Somewhat inefficient memory utilization as the RAM occupancy reached in the range of 48% | • Reasonably efficient memory utilization | • Best memory utilization that remains almost constant with scaling clients | • Least memory utilization efficiency. Memory utilization grows with scaling client count. |

### 6.2.1 Flower FL Framework

The results presented in Table 2 and the preceding figures show that the overall training process in the Flower framework exhibits variability with different client counts, with a minor decrease in total training time as the number of clients increases. The loss tends to increase when training using a higher client count, with the Flower platform displaying a higher total loss in comparison to the alternative platforms across all client numbers. Despite the rise in loss, accuracy remains relatively stable, experiencing only a small decline as the number of rounds increases. Notably, both CPU and RAM usage are higher than the other frameworks.

### 6.2.2 FedML Framework

In comparison, it can be observed that the total training time for FedML is consistently lower in comparison with Flower. Furthermore, as the number of clients increases, there is an observable trend of decreasing loss and improving accuracy, but with some observed fluctuations. Notably, although the loss is generally lower than that of Flower, the accuracy in the Flower experiments remains significantly higher than in FedML. CPU and RAM usage remain relatively stable for FedML; however, there is a marginal increase in CPU usage as the number of clients grows. As can be observed from Figure 9, CPU usage for FedML rises during the training process and oscillates at high values whereas the RAM usage remains stable at a relatively low value.

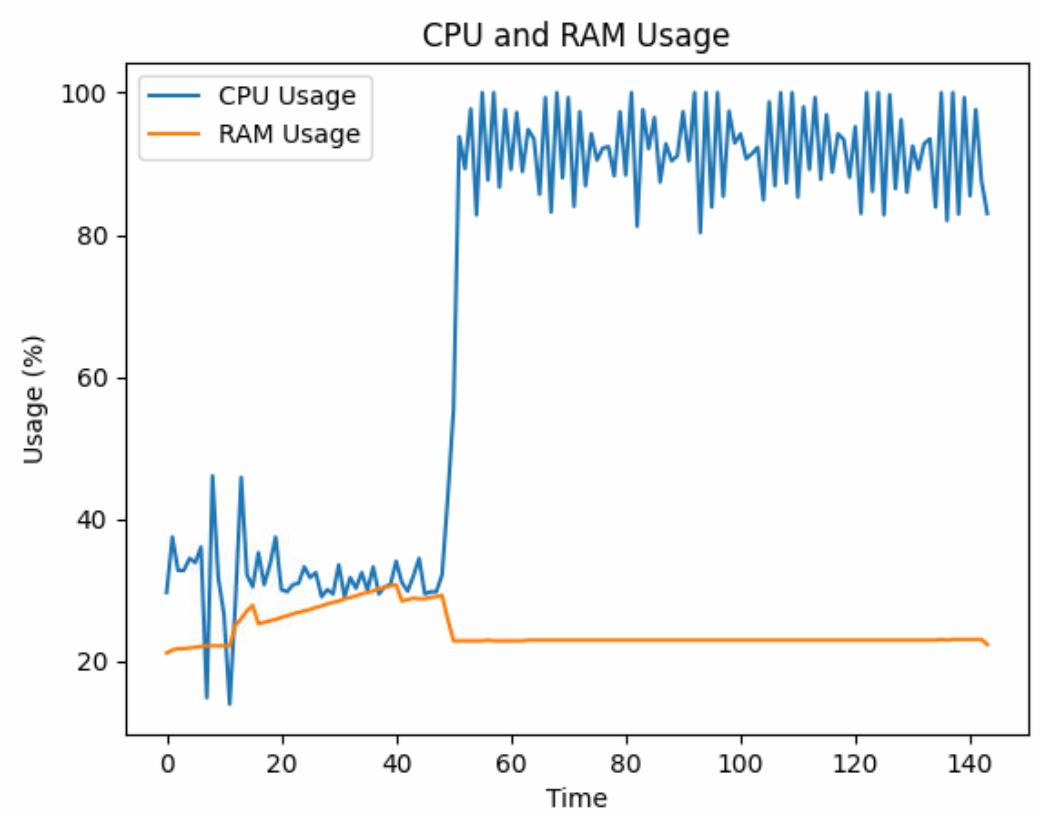

Figure 9 – CPU and RAM Usage Patterns for the FedML Framework.

### 6.2.3 Substra Framework:

The Substra framework demonstrated a consistent and notable rise in the total training time as the number clients is increased, surpassing the total training time of both FedML and Flower by large margins. Despite variations in loss across different client counts, the accuracy value remains consistently high. As can be observed in Figure 10, the loss and accuracy metrics for Substra improve as the training process progresses through the different rounds.

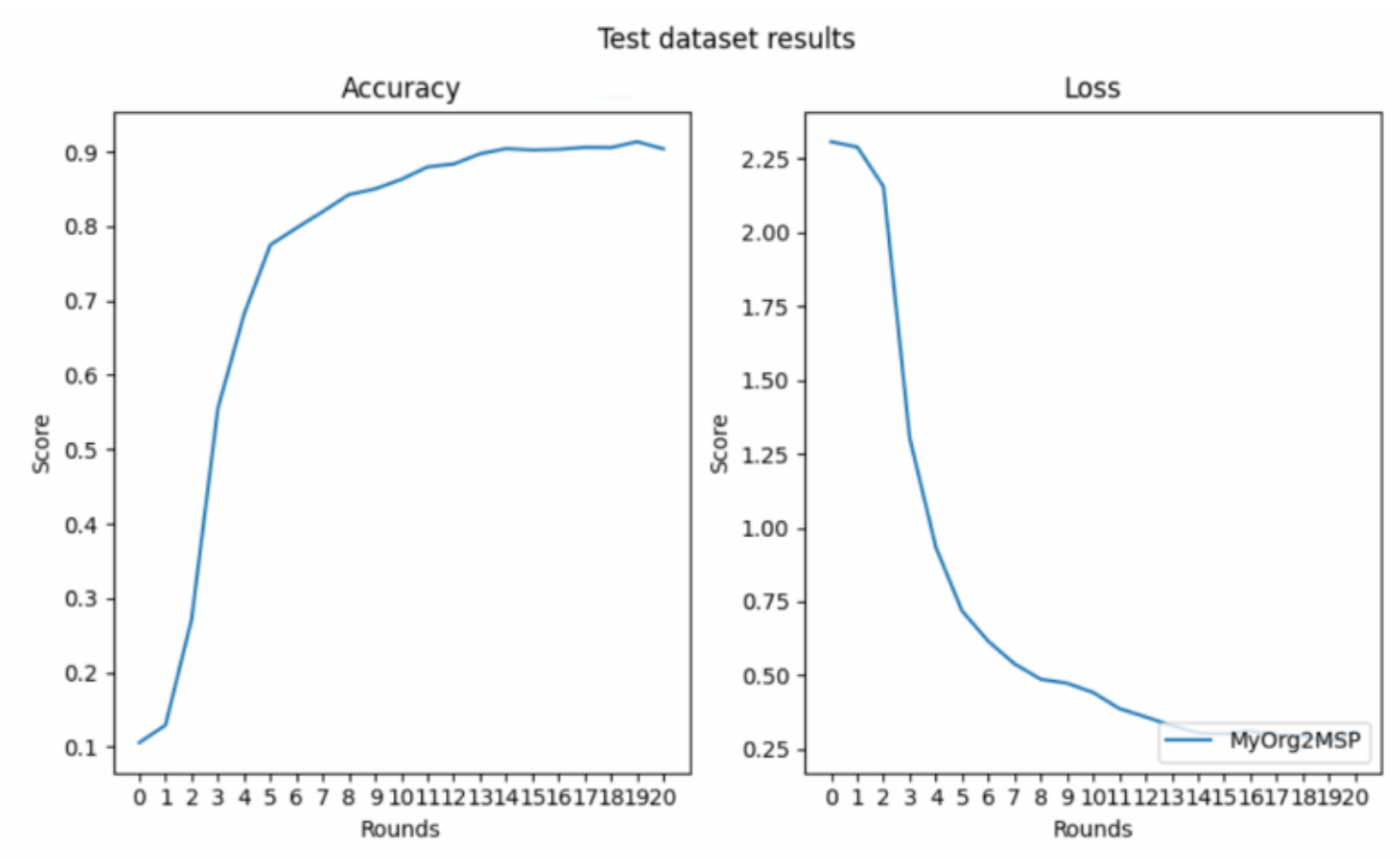

Figure 10 – Substra Accuracy and Loss Graphs using 2 Clients.

Interestingly, as the number of clients grows, there is a slight decrease in CPU usage, while RAM usage remains relatively stable throughout the experiments. Because of its computation efficiency and low memory utilization, Substra seems to be the most appropriate for resource-limited implementations such as IoT and edge-computing hardware. However, it suffers from a significant scalability issue with respect to training time.

### 6.2.4 OpenFL framework:

The data show that as the number of clients is increased, there is a corresponding moderate increase in total training time for the OpenFL framework. Notably, this increase is not as substantial as the increase exhibited by Substra. Throughout the evaluations, both loss and accuracy of OpenFL remain stable, with only minor fluctuations observed across different client counts. Interestingly, as the number of clients increases, RAM usage rises, while CPU usage remains relatively stable with only minor fluctuations. These experiments suggest that while there is a notable impact on training time and resource utilization, the model's performance, as measured by loss and accuracy, remains generally robust and consistent across the different number of clients involved in the training process. OpenFL appears to be the most scalable of all platforms, as it displayed consistent accuracy, loss, and training time across different client counts and stable CPU and RAM usage with minor variations.

# 7 Conclusion

Federated Learning (FL) represents a paradigm shift in collaborative machine learning, emphasizing privacy and decentralized data processing. This research aims to evaluate the performance of prominent FL frameworks - Flower, FedML, Substra, and OpenFL. Their performance is evaluated based on experiments that measure the effect of scaling client counts on metrics such as total training time, loss and accuracy values, and CPU and RAM usage during training.

The experimental results show that Flower achieves the best accuracy value (98% on average) compared to the other framework, albeit with an unusually high loss value regardless of the client count. CPU and RAM usage for Flower remain relatively stable, with a marginal increase in CPU usage as the number of clients grows. On the other hand, Flower seems to exhibit a significant inherent training overhead since its training time is consistent, but comparatively high compared to the other frameworks. The FedML framework achieved a very low accuracy level of 68% on average, which became slightly better with increased client count. Its total training time was amongst the lowest, growing linearly with increasing client count. FedML exhibited a very significant computational overhead as it consistently consumed in excess of 95% of CPU cycles. This would not bode well for implementation in resource-limited applications.

Substra provided the best efficiency in terms of resource utilization, placing a very low burden on both the CPU and the RAM usage, regardless of client count. It also achieved consistently good loss and accuracy measures. On the other hand, Substra presented a significant issue with exponentially increasing total training time as the number of clients is increased. OpenFL presented the most compelling framework as it consistently achieved very good loss and accuracy measures (above 96% on average) and reasonably stable resource utilization across the different client counts. While its total training time was somewhat elevated for low client counts, it remained substantially stable as the number of clients increased.